\documentclass[12pt]{article}
\usepackage{graphicx}
\usepackage{a4wide}

\begin{document}

\title{Direct measurements of the effects of salt and surfactant on interaction forces between colloidal particles at water-oil interfaces}

\author{Bum Jun Park, John P.\ Pantina, and Eric M.\ Furst\thanks{furst@udel.edu}\\ Department of Chemical Engineering,
University of Delaware\\ Newark, Delaware 19716, USA\\[0.5cm]
 Martin Oettel\\
Institut f$\ddot{\mathrm{u}}$r Physik, Johannes-Gutenberg-Universit$\ddot{{\mathrm{a}}}$t Mainz \\
 WA 331 55099 Mainz, Germany\\[0.5cm]
Sven Reynaert and Jan Vermant\thanks{jan.vermant@cit.kuleuven.be}\\ Department of Chemical
Engineering, Katholieke Universiteit Leuven\\ W.\ de Croylaan 46, B-3001 Leuven, Belgium}
\date{}

\maketitle

\newpage
\setlength{\baselineskip}{24pt}

\begin{abstract}
The forces between colloidal particles at a decane-water interface, in the presence of low
concentrations of a monovalent salt (NaCl) and of the surfactant sodium dodecylsulfate (SDS) in
the aqueous subphase, have been studied using laser tweezers. In the absence of electrolyte and
surfactant, particle interactions exhibit a long-range repulsion, yet the variation of the
interaction for different particle pairs is found to be considerable. Averaging over several
particle pairs was hence found to be necessary to obtain reliable assessment of the effects of
salt and surfactant. It has previously been suggested that the repulsion is
consistent with electrostatic interactions between a small number of dissociated charges in the
oil phase, leading to a decay with distance to the power -4 and an absence of any effect of
electrolyte concentration. However, the present work demonstrates that increasing the electrolyte
concentration does yield, on average, a reduction of the magnitude of the interaction force with
electrolyte concentration. This implies that charges on the water side also contribute
significantly to the electrostatic interactions. An increase in the concentration of SDS leads to
a similar decrease of the interaction force. Moreover the repulsion at fixed SDS concentrations
decreases over longer times. Finally, measurements of three-body interactions provide insight into
the anisotropic nature of the interactions. The unique time-dependent and anisotropic interactions
between particles at the oil-water interface allow tailoring of the aggregation kinetics and
structure of the suspension structure.

\end{abstract}

\newpage
 \setlength{\baselineskip}{24pt}

\section*{Introduction}
The behavior of colloidal particles at liquid interfaces is of great practical importance [see
ref. \cite{BOOKBinks} for a recent review]. For example, particles stabilized emulsions and foams
can be produced \cite{Ramsden,Pickering,Aveyard}, which are encountered in both established and
emerging areas of technology. Examples include froth flotation or ice-cream production, as well
colloidal self-assembly and the production of colloidal capsules (colloidosomes) \cite{Weitz}.
Exploiting properties of particles at interfaces can also be used to create crystalline assemblies
of varying form \cite{VELEV} or colloidal microcrystals of well defined geometry \cite{pine}. As
for all applications where colloidal materials are being processed, there is a need to understand
the colloidal interactions in order to have control over the structure, and hence the properties.
For example, the most efficient stabilization of Pickering emulsions will probably be achieved by
a relatively low surface coverage and weakly aggregated structure at the droplet interfaces
\cite{Vignati,Midmore}. Planar monolayers of particles with controlled interactions also enable
fundamental studies of colloid physics, such as crystal melting \cite{Maret}, aggregation
\cite{Earnshaw1,Earnshaw2,Hansen,Reynaert}, packing and self-assembly of non-spherical particles
\cite{Loudet,BAsa} and even elucidate the effects of shear or extensional flow on the
microstructure of suspensions \cite{Stancik,Hoekstra}. In all cases, predictive control over the
generated structures is needed, necessitating a detailed understanding of the colloidal
interactions.\\

Despite the thorough understanding of colloid and surface forces developed over the past century
\cite{Israelachvili,Russel}, the presence of an interface adds significant complexity to the
colloidal interactions. A first illustration hereof is the remarkable stability of two-dimensional
colloidal crystals when the ionic strength of the aqueous phase is increased
\cite{Earnshaw2,Aveyard200a,Reynaert}. Monolayers of initially charged polystyrene latex beads
spread on an aqueous subphase containing as much as 0.5 M of a monovalent or divalent salt were
observed to be stable for several days before aggregation set in \cite{Earnshaw2,Reynaert},
whereas the bulk system aggregates on the timescale of minutes. Secondly, there is also a strong
dependence on how monolayers are prepared \cite{Spanjaardenhfdstk}. For example, when particles are
first spread onto a virgin interface, and salt is added to the aqueous subphase slowly after that,
the aggregation proceeds as expected, after the long induction process mentioned. Yet, particles
spread onto an aqueous subphase already containing the salt can result in a monolayer where
crystalline patches and dense aggregates coexist \cite{Reynaert}. Finally, probably the most
remarkable illustration of the complexity of the interaction forces is the occurrence of so-called
mesoscale structures. At low surface coverage, fascinating patterns can be readily observed,
including voids, line patterns, soap froths and particle loops
\cite{RuizGarcia1,RuizGarcia2,Ghezzi1,Ghezzi2}. They are consistent with the presence of an
attractive, secondary minimum in the intercolloidal potential at distances spanning several
particle diameters, and with a well depth on the order of thermal energy $kT$, although it has
also been suggested that they are a consequence of like-charge attractions \cite{TONG}.\\

There appear to be two main causes for this increased complexity. First, the repulsive
electrostatic interaction occurs through two media with different values of the dielectric
constants. Second, the presence of charged particles can create distortions of the interface, which
can give rise to (predominantly attractive) lateral capillary interactions. We will briefly review
the understanding of both classes of interactions.\\

Compared to similar particles in a bulk liquid, the electrostatic repulsion between charged
particles is enhanced at a water-low dielectric medium interface, as was recognized early on by
Pieranski \cite{Pieranski}. At the interface, a particle has an asymmetric counterion distribution
which results in a dipole-dipole interaction through the phase of low dielectric constant. It has
been shown that the corresponding interaction is repulsive and long-ranged. The force decays
inversely with the fourth power of the separation between the dipoles, and depends strongly on
electrolyte concentration, through the Debye screening length $\kappa^{-1}$ \cite{HURD,MONCHO},
\begin{equation}
F_{E} = \frac{3 \varepsilon_{oil} }{ 2\pi \varepsilon_{0} \varepsilon_{W}^2  } \cdot
\frac{q^2 \kappa^{-2}}{r^4} \label{eq:ESHurd}  \qquad.
\end{equation}
The permittivity of free space and the relative dielectric constant of water and oil are given by
$\varepsilon_0$, $\varepsilon_{W}$ and $\varepsilon_{oil}$, respectively.  Because only the
particle surface charges within a distance $\kappa^{-1}$ from the three-phase contact line
contribute to the equivalent point charge $q$, Aveyard et al. suggested that the force depends on
$\kappa^{-4}$ \cite{Aveyardprl}. The relative independence of two-dimensional colloidal crystals
to the electrolyte concentration which is experimentally observed \cite{Earnshaw2,Aveyard200a},
suggests that the Hurd model does not adequately describe the electrostatic interaction. A recent
explanation for a weaker dependence on $\kappa^{-1}$ is that charge renormalization is important
due to the high surface charge density \cite{OettelCR}. This problem has been studied by solving
Maxwell's equation in the oil phase and  the Poisson-Boltzmann equation in the water phase,
assuming constant charge density $\sigma$ on the colloid surface exposed to water. The dipole
character of the interaction is retained, and far enough from the particle surface one recovers
eq.~\ref{eq:ESHurd}, but  $q$ needs to be replaced by an effective charge $q_{eff} =q\,g$. For a
particle of radius $R$, the charge renormalization function $g$ depends primarily on a reduced
inverse screening length $\kappa^*=\kappa R$, a reduced charge density $\sigma^*=\sigma (e R)/(k
T\,\varepsilon_0 \varepsilon_W)$ and, to a minor extent, also on the ratio of the dielectric
constants between the different media and on the geometry through the contact angle $\theta$ ($e$
is the elementary charge). For sufficiently high surface charge densities the scaling of $F_E$
with the screening length becomes \cite{OettelCR}:
\begin{equation}
F_{E} \propto \ln^2 \frac{\sigma^*}{\kappa^*}
\label{eq:Oettel}
\end{equation}
which yields a weak power--law dependency of  $~ \kappa^{-0.8 \cdots -0.4}$ for physical charge
densities $\sigma=1 \dots 10$ $\mu$C/cm$^2$ ($\sigma^*\approx 800\dots 8000$), which clearly is
different from the linear results.\\

Alternatively, it has been suggested that 'monopolar' Coulomb repulsive interactions must also be
taken into account. Aveyard {\it et al.}\ \cite{Aveyard200a} hypothesized that a small number of
unscreened surface electric charges, possibly arising from dissociated surface groups, are
stabilized by water trapped on the rough particle surface. The force acting between the particles
can be calculated as a function of the interparticle distance, using Stillingers approach
\cite{Aveyard200a,Spanjaardenhfdstk}, where the charges in the oil phase are represented by a
point charge $q_{oil}$ located at a distance above the oil-water interface. For sufficiently large
particle separations, this force (labeled $F_{Eo}$) has the characteristic dependence of a
dipole-dipole interaction on interparticle distance ($r$) \cite{Aveyard200a,Aveyardprl}:
\begin{equation}
F_{Eo} \approx \frac{q_{oil}^2}{4\pi \varepsilon_{oil} \varepsilon_0} \left [{\frac{1}{r^2} -
\frac{r}{( 4 \zeta^2 + r^2)^{3/2}}} \right ]  \qquad , \label{eq:ES}
\end{equation}
with $\zeta$ being the distance of an equivalent point charge over the dimensions of the particle,
\begin{equation}
\zeta = R \frac{(3 - \cos \theta)}{2} \label{eq:ES2} \qquad,
\end{equation}
where $\theta$ is the contact angle and $R$ the particle radius. The charge $q_{oil}$ is given by
the product of the particle surface area immersed in the oil, the surface charge density $\sigma$
and the fractional degree of dissociation, $\alpha_{oil}$, of the ionizable groups at the
oil-particle interface:
\begin{equation}
q_{oil} = 2 \pi R^2 (1- \cos \theta) \cdot  \sigma \cdot \alpha_{oil} \qquad.
\end{equation}
The expected force contribution from this `oil side' effect is independent of $\kappa$. The limit
for  ($r/ \zeta )^2 \gg 1$ of Eqn.\ref{eq:ES} is
  \cite{Aveyard200a,Aveyardprl}
\begin{equation}
F_{Eo} \sim \frac{3q^2}{8 \pi \varepsilon_{oil}\varepsilon_0}\frac{\zeta^2}{r^4}.
\end{equation}
\\

Direct measurements using optical tweezers confirmed the $r^{-4}$ dependence, and it was shown
that the measured force did not change significantly when 0.1mM of NaCl was added to the aqueous
phase. However, some issues remain, as the data set is rather limited and the inferred degree of
dissociation that produces $q_{oil}$, ranges from 0.033\% \cite{Aveyardprl} to 1\%
\cite{Aveyard200a}, for the same type of particles, depending on the measurement technique, i.e.
pair interaction measurements using laser tweezers versus bulk surface pressure isotherms.
Molecular dynamics simulations suggest that the combination of the short-range dipole-dipole
interactions and long-range charge-charge repulsions adequately describes the experimentally
observed isotherms, provided $\alpha$ equals 0.4\% \cite{sunstirner}, which is an order of
magnitude above the results from the direct tweezer measurements.\\

Considering the  attractive contributions, a dominant feature of colloidal suspensions at
interfaces is the occurrence of lateral capillary forces. They appear when the presence of
particles at a fluid phase boundary causes perturbations in the interfacial shape, which subsequently overlap \cite{Krachlrev}. For particles which cause similar
interfacial deformations, i.e. both downward or both upward, the lateral capillary forces are
attractive in nature. Assuming a superposition of the interfacial deformation profiles,
Kralchevsky and Nagayama \cite{Krachl1} calculated the interaction force when gravity is the
determining force. For the particles studied in the present work, these gravitational forces are
small (${\cal O}[10^{-11}$N]). More importantly, the dipolar electric field of the particles can
also create a local force imbalance due to the Maxwell stress tensor, which can lead to
electro-capillary interactions and medium ranged flotation-like forces \cite{Oettel1,Danov}. The
electro-capillary force scales like
\begin{equation}
\label{eq:Fcap}
F_{cap} \sim  - F_{E} \epsilon_F \qquad .
\end{equation}
Hence, it is influenced by the same factors as the electrostatic force, but has an additional
prefactor $\epsilon_F$, which is given by the ratio of the total electrostatic force acting on the
colloid and the surface tension force scale $\gamma R$  \cite{Oettel1}. The magnitude of the
electro-capillary force hence depends on the ratio of the total (vertical) electrostatic force
acting on the colloid to $\gamma R$. Finally, capillary interactions depend on the shape of the
meniscus, aspects such as asperities on the particle surface, particle or agglomerate shape. The
interaction can be repulsive or attractive depending on the relative orientations of the interface
deformations. This also implies that when doublets or irregular aggregates are formed with
possibly complex undulating contact lines, the resulting capillary interactions will no longer be
isotropic \cite{Krachlrev,Stamou,Vandenende}.\\

 The aim of the present work is to investigate the control of colloidal interactions by
the addition of small amounts of surfactant or salt to the aqueous subphase. Detailed measurements
of the effect of electrolyte concentration should shed light on the importance of contributions of
the charges on the oil and water sides to the electrostatic interaction force. The study of the
effect of changing surfactant concentration is motivated by earlier observations that small
amounts of either anionic or cationic surfactants lead to an efficient destabilization of the
monolayer \cite{AveyardLangm,Reynaert}. Adding ionic surfactants will affect the screening length,
the interfacial tension, but also the wetting conditions. In the present work, direct measurements
of the effect on the overall interaction force are presented. The time-dependence of the
inetercation force as well as the spatial anisotropy of the capillary interactions will be
investigated and its role in structure formation will be measured in detail.

\section*{Materials and Methods}
An oil-water interface was created using n-decane (Acros Organics, 99+\%) and deionized water.
Prior to use, polar components were removed from the decane using an adsorption onto
aluminiumoxide powder (Acros Chemical, acidic activated, particle size 100-500$\mu$m). To modify
the wetting properties and dielectric properties of the aqueous subphase small amounts of an ionic
surfactant, sodium dodecylsuplhate (SDS, Sigma Aldrich 98+\%), sometimes in combination with a
monovalent salt (NaCl), were used. Monodisperse, charge stabilized polystyrene particles (9.1
$\mu$C/cm$^2$ and 3.1$\pm$0.2$\mu$m diameter) were obtained from Interfacial Dynamics Corporation.
The charge is a result of the presence of sulfate groups at the particle surface.\\

Particle monolayers at the oil water interface are prepared according to the procedure outlined in
Reynaert et al.\ \cite{Reynaert}. A stable colloid monolayer is first prepared, and subsequently,
salt or surfactant are carefully added to the aqueous subphase, without going through the oil
phase, in a specially designed fluid cell to accommodate the short working distance of the
microscope objective (i.e. a 200$\mu$m working distance for the 40$\times$ water immersion
objective) required in the optical tweezer experiments.\\

\begin{figure}[ht!]
\begin{center}
\includegraphics[width=12cm]{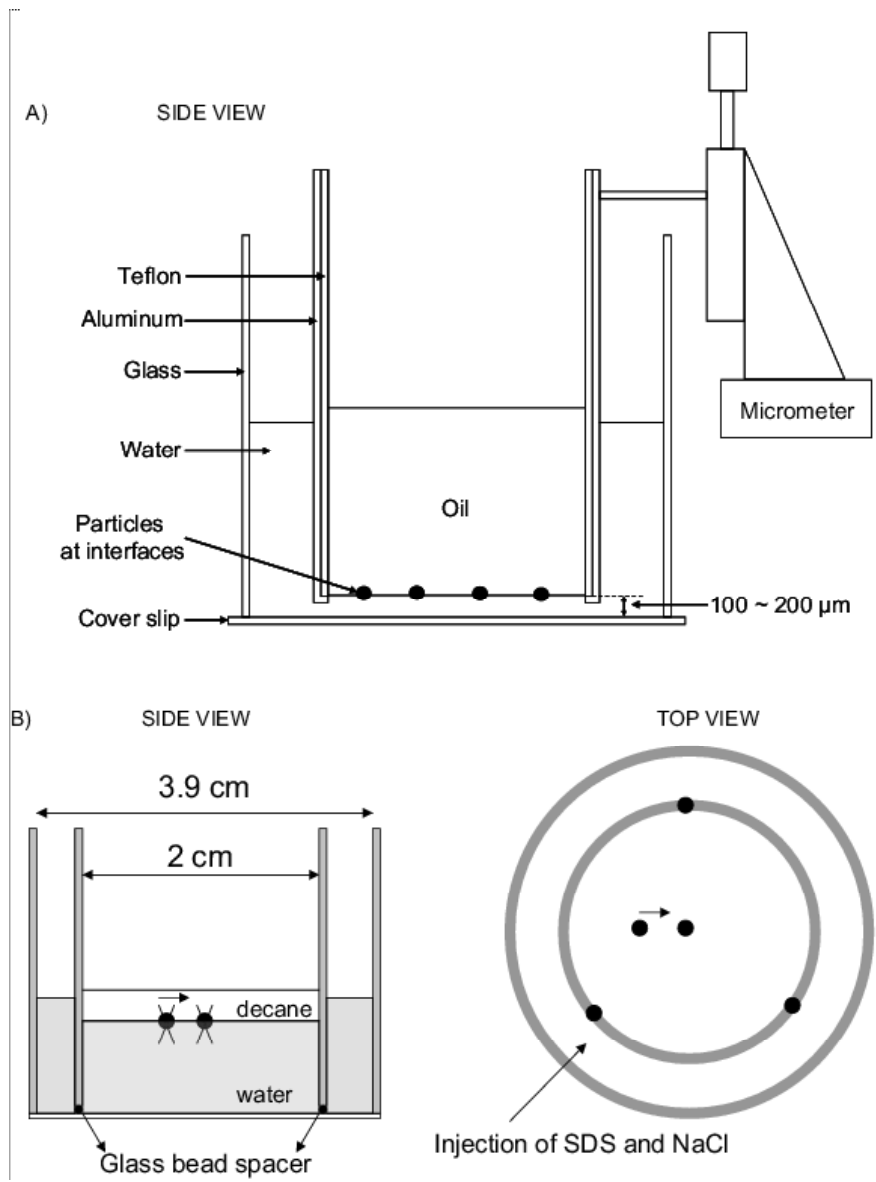} {\caption{\small Experimental cells used for the optical
tweezer experiments (a) Aluminium cell (b) Glass cell. See text.} \label{fig:cell}}
\end{center}
\end{figure}

Two experimental cells were used, as shown in Figure~\ref{fig:cell}.  The first experimental cell
in fig.~\ref{fig:cell}.a consists of an inner and outer cylinder made of aluminum.  The inner
cylinder is held by a micrometer, while the outer cylinder is attached to a 40mm circular No. 1.5
coverglass (Fisherbrand) using a fast curing UV epoxy (Norland Products, NOA 81). A teflon ring is
inserted into the bottom of the inner cylinder in order to pin the contact line of the oil-water
interface. Decane is placed in the cell first, achieving hydrostatic equilibrium between the inner
and outer cylinder, and water is added to the inner cylinder.  The height of the inner cylinder is
adjusted to by raising or lowering it with the micrometer.  Using the height adjustment along with
the addition or removal of decane and
water, ensures a flat interface.\\

The second experimental cell, as shown in Figure~\ref{fig:cell}.b is built using two concentric
glass rings (height of 1cm, outer diameter 39mm and 18mm, respectively, and a wall thickness of 1
mm), confining the oil-water interface within the inner ring. Again, the rings are attached to a
40mm circular No. 1.5 coverglass (Fisherbrand) using a fast curing UV epoxy (Norland Products, NOA
81). The outer ring is completely sealed to prevent any fluid leaks; however, a three-point
attachment using 200$\mu$m glass beads as spacers between the inner ring and the coverglass are
used. This enables the system to achieve hydrostatic equilibrium, thus providing control over the
water layer by extracting water from or adding water to the outer ring. In both experimental
setups, all glassware is fire treated using a propane torch immediately before constructing the
cell. This reduces the water-glass contact angle to below 15$^{\circ}$, achieving good wetting
conditions for the water
layer.\\

The optical tweezer setup used is described in greater detail in Pantina and Furst \cite{Pantina}.
Two methods were used in this study to measure forces with the optical traps. To calibrate the
trapping force of the tweezers, particles were held at the interface and subjected to drag forces
by translating the microscope stage at constant velocities, $U$.  The displacement of the particle
from the center of the optical trap is measured as a function of the Stokes drag force $F_S = 6\pi
a \eta_{\mathrm{eff}} U$, where the effective viscosity depends on both phases,
$\eta_{\mathrm{eff}} = [ \eta_{oil}(1-\cos \theta) + \eta_{water}(1+\cos \theta)]/2$.  This
provides a calibration of the trap force profile and maximum trapping force.\\

An alternative method was used in some cases, in which the particle response to a sinusoidal
oscillation of the optical trap is characterized.  This method can better adapt to situations in
which the curvature of the oil-water meniscus is significant, and dragging the particle at
constant velocity is not feasible. At low frequencies, the particle is trapped and follows the
beam. When the frequency of the oscillation is increased, the increasing viscous drag force acting
on the particle causes it to leave the trap. Using the calculations from Faucheux et al.\
\cite{Faucheux}  the calibration constants can be obtained from the frequency at which the system
transitions from ``phase lock'' to ``phase slip''. In all experiments, it was verified that
measurements of the interaction forces where independent of the laser intensity. This implies that
the optical tweezers are not exerting a net force on the particles. If a possible vertical net
force were strong enough, attractive long-ranged forces should appear in the force
measurements \cite{Oettel1}.\\

\begin{figure}[ht!]
\begin{center}
\includegraphics[width=13cm]{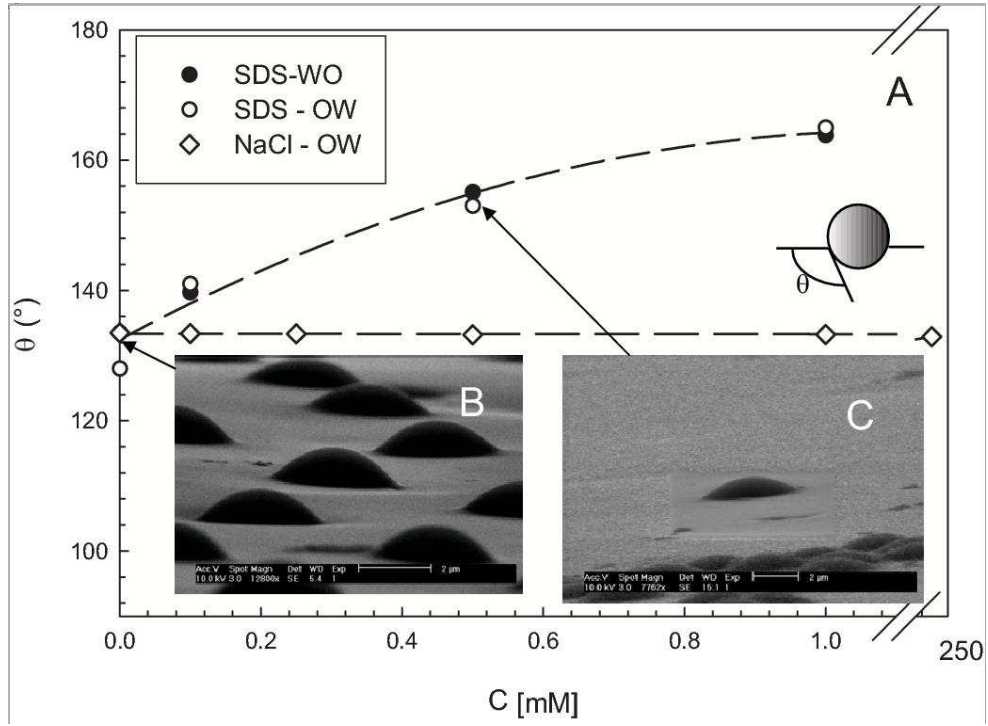} {\caption{\small Effect of small amounts SDS and NaCl concentration on the contact angle on a
polystyrene cast film for water-oil (WO) and oil-water (OW) interfaces (B) SEM image of gel
trapped particles originally at a water-decane interface.
 (C)  SEM image of a gel particle trapped originally at an oil-(water+0.5mM SDS) interface.}
\label{fig:contact}}
\end{center}
\end{figure}

The contact angle $\theta$ of the particles at the oil water interface is measured using two
independent methods. First, a film is cast of the PS particles dissolved in chloroform. A water
droplet is subsequently gently placed on the PS film, and a decane environment is then generated
around the droplet. A goniometer (CAM 200, KSV instruments) is used to measure $\theta$. To check
for contact line hysteresis, some measurements were also performed on the reverse system of a
decane droplet in a water environment. The results (from ref. \cite{Reynaert}) are given in figure
\ref{fig:contact}. The contact angle increases as SDS is added, i.e.\ particles are pushed into
the oil-phase. Becasue casting of the PS films may mead to slight differences in surface chemistry
and changes in the wetting properties, we also used the so-called gel trapping technique of Paunov
\cite{Paunov} to verify our results. The particles are spread at the oil-water surface and the
water phase is gelled using a nonadsorbing polysaccharide. The particle monolayer trapped on the
surface of the gel is then replicated using a PDMS-elastomer. The particles, now embedded within
the PDMS surface, are imaged with high resolution SEM (Philips XL300FEG-ESEM). Fitting the contour
of the particles to a sphere gives information on the position and the particle contact angle at
the air-water or the oil-water interface. Typical micrographs for a water-decane interface are
given in the insert of figure \ref{fig:contact}(B). The SEM image for a subphase containing 0.5mM
SDS is shown in figure \ref{fig:contact}(C). The reduction of the exposed particle surface is
consistent with the particle being pushed into the oil phase. The value of the contact angle can
be obtained by measuring the height above the interface and dividing it by the radius at the plane
of intersection. The value obtained for the particle at the water- decane interphase is  118 $\pm$
2 $^{\circ}$ and for the system water( + 0,5 mM SDS)- decane : 142 $\pm$ 2 $^{\circ}$. These
values are consistent with those observed from the cast films, the small differences being
attributed to the presence of gellan in the aqueous phase. As expected in this range of
concentrations, adding salt has a only a very limited effect on the contact angle, a small
decrease is observed in the case film experiments, on the order of one degree, at the highest
concentration studied (250mM).

\section*{Results and Discussion}

\subsection*{Pair interaction}
\subsubsection*{Effect of NaCl}
 Two particles are trapped at a sufficiently wide initial separation such that the repulsive
interaction between the particles is weak. One particle is then moved step-wise to smaller
separations.  The displacement of the stationary particle is monitored in its optical trap.  Using
the calibrated trap stiffness, the force at a number of separations can be easily found in a
manner analogous to the spring displacement in an AFM or surface forces apparatus.\\

\begin{figure}[ht!]
\begin{center}
\includegraphics[width=13cm]{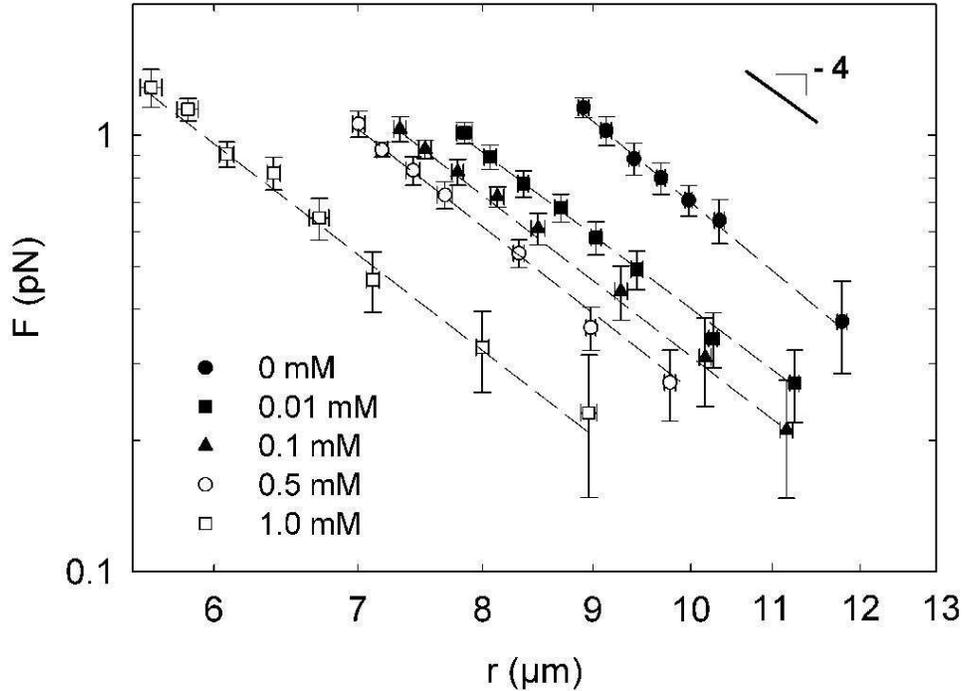}
{\caption{\small Measurement of the pair interaction between PS
particles at the decane-water interface for different
concentrations of NaCl in the subphase, approximately one hour
after the monolayer was prepared.} \label{fig:SALTinitial}}
\end{center}
\end{figure}

Figure \ref{fig:SALTinitial} shows the force profile measured for the pure water-decane interface
and at four different NaCl concentrations, one hour after preparing the monolayer. Four aspects
can be deduced. In agreement with the earlier observations of Aveyard et al \cite{Aveyardprl}, the
force is repulsive, and is observed to scale as $r^{-4}$. Secondly, the force is found to be
small, even somewhat smaller than the force values obtained by Aveyard et al. \cite{Aveyardprl}
using tweezers, but especially more than three orders of magnitude smaller than the values
inferred from surface pressure area isotherms using a Wilhelmy plate and balance
\cite{AveyardLangm}. If the interaction as measured by the tweezer would be attributed to charges
in the oil phase, this would correspond to a degree of dissociation smaller than 0.01\%. The
tweezer experiments were carefully checked for artefacts, using different particle arrangements,
and even utilizing multiple particles, e.g. placed on a hexagon, with a particle in the center.
Nonetheless, on average, consistent values were always obtained. Most probably, the indirect
Wilhelmy technique is less reliable for the particle laden interfaces. The Wilhelmy technique
assumes that only a monolayer of particles, with the same structural arrangements as at the
interface, stick to the plate and transmit the force. Changes in both the interaction force and
the local structural arrangements may be caused by particle adsorption onto the plate, which may
be complicated by the local curvature of the interface. Gel trapping measurements revealed
slightly denser packed structures close to the Wilhelmy plate, and a corresponding enhancement of
the electrostatic force is subsequently measured by the plate. This suggests that interpreting
pressure-area isotherms obtained with the Wilhelmy technique in terms of the pair interaction
should be handled with care. A third observation from the data in figure \ref{fig:SALTinitial} is
that the dependence on particle separation is also conserved for the systems containing salt in
the subphase. And finally, it is shown that increasing the electrolyte concentration in the
subphase significantly decreases the magnitude of the repulsive interactions. This is true {\sl on
average}; however, significant heterogeneity in the pair interactions are observed, depending on
the particle pairs used in the force measurements, as discussed below. The error bars mainly
reflect this variation on the measured interaction force.\\

\begin{figure}[ht!]
\begin{center}
\includegraphics[width=13cm]{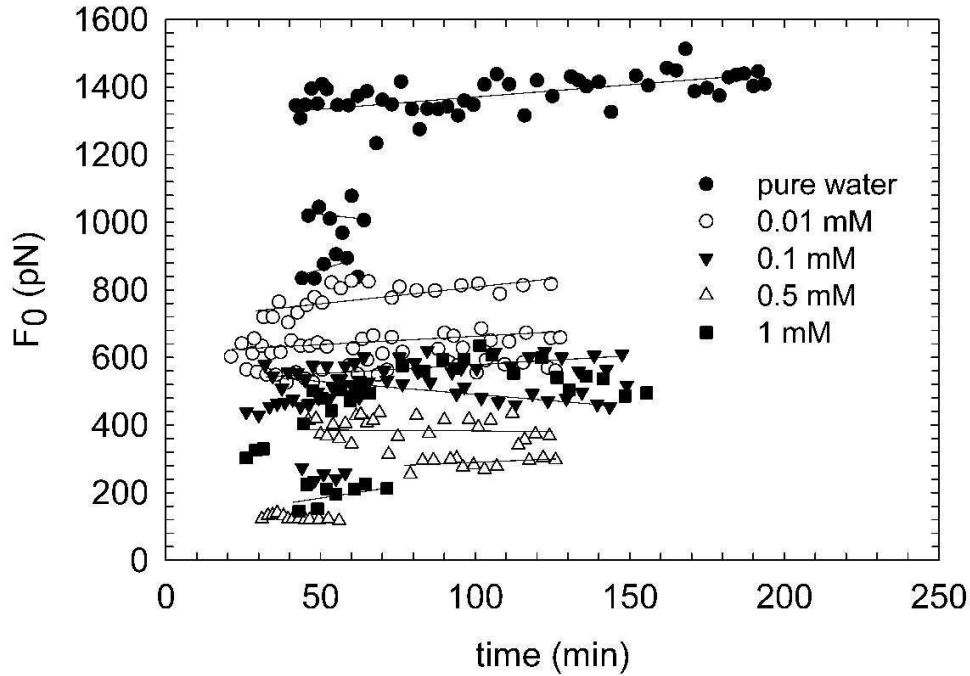}
{\caption{\small Typical measurements of the force between PS
particles at the decane-water interface for pure water and three
concentrations of NaCl in the subphase.
 $F_0$ is the amplitude of the  dipole force parametrization  $F(r)=F_0\,(R/r)^4$.} \label{fig:SALT1}}
\end{center}
\end{figure}

To evaluate the presence of any time dependence, the rate of change in the repulsive particle
interaction is measured. Figure \ref{fig:SALT1} shows the force for several particle pairs at
three different concentrations. For times up to three hours, no measurable change occurrs in the
interaction force, either for the pure system, or for the systems containing salt in the subphase.
Note, however, that the pair interactions exhibit significant variability under the same
conditions. The force differs by as much as a factor 2. This could be due to a local surface
charge inhomogeneity at the particle surfaces which could cause variations in the repulsive part
\cite{TONG}, or due to effects of nanoscale particle roughness or effects of the charge
inhomogeneity on the homogeneity of the wetting which would lead to an attractive force of
considerable magnitude \cite{Stamou}.

\subsubsection*{Effect of SDS}

\begin{figure}[ht!]
\begin{center}
\includegraphics[width=13cm]{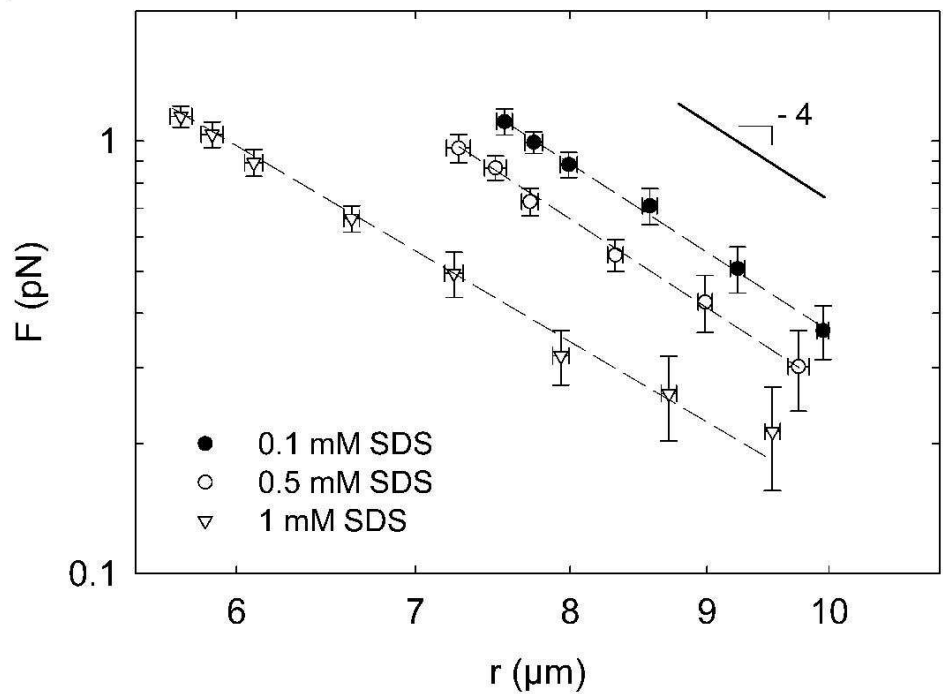}
{\caption{\small Measurement of the pair interaction between PS
particles at the decane-water interface for different
concentrations of SDS in the subphase, approximately one hour
after the monolayer was prepared.} \label{fig:SDSinitial}}
\end{center}
\end{figure}

SDS was added at concentrations well below its CMC to the subphase.  Apart from changing the
screening length in water by acting as an electrolyte, SDS also changes the wetting properties,
and the particle is pushed further into the oil phase due to the change in the contact angle, as
demonstrated in figure~\ref{fig:contact}. When SDS is slowly added to the subphase and one assumes
the charges on the oil side to stay the same, the effect of surfactant on the interaction forces
should yield insight into the question of whether charges on the oil side or on the water side
dominate the repulsive forces. When the electrostatic forces on the oil side dominate, the
interaction force should increase, the particles are pushed further in the oil phase
(fig.\ref{fig:contact}), increasing both the area exposed to the oil and the height above the
interface. A decrease is expected when the water-side electrostatic forces dominate as the exposed
area is decreased and the screening length increases because of the ionic nature of the
surfactant.  Figure \ref{fig:SDSinitial} shows the force profile measured at three different SDS
concentrations approximately one hour after preparing the monolayer. Similar to the addition of
small amounts of NaCl, increasing the SDS concentration in the subphase decreases the magnitude of
the repulsive interactions. Again, this is true only on average, as significant heterogeneity is
also present here. As the repulsive interaction decreases at higher SDS concentrations, the
repulsion softens, shown by the slight decrease in the scaling exponent with particle separation.
This suggests that a long-range attraction begins to play a role as particles come into closer separations.\\

\begin{figure}[ht!]
\begin{center}
\includegraphics[width=13cm]{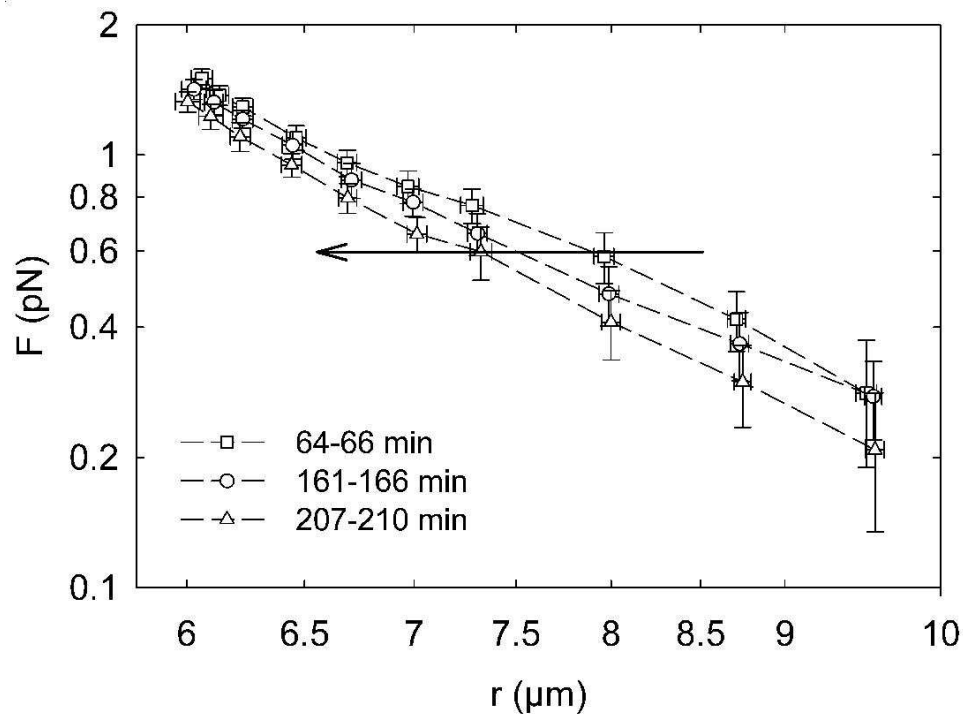}
 {\caption{\small Comparison of the interaction force for
a subphase containing 0.1 mM SDS at different times after preparation of the monolayer.
\label{fig:SDStime1}}}

\end{center}
\end{figure}

Whereas the results above indicate that initial electrostatic repulsion remains significant,
earlier studies showed that surfactants can completely  {\sl de}stabilize an initially crystalline
monolayer \cite{AveyardLangm,Reynaert}, albeit sometimes after very long incubation times.
Therefore, the time dependence of the pair interaction was also investigated. Using a similar approach
as in figure~\ref{fig:SDSinitial}, the interaction force as a function of separation was monitored
at different time intervals after preparation of the monolayer, as shown in
figure~\ref{fig:SDStime1}. It is observed that the repulsion between the particles becomes weaker
with time.\\

To examine the time dependence further, the rate of change in the repulsive particle interaction
was measured.  Figure \ref{fig:SDStime1} shows the force between particle pairs for 0.1mM SDS at
three time intervals over the span of one to 3.5 hours.  Over this period, the repulsion is
observed to decrease. Note, however, that particle pairs again exhibit significant variability
under the same conditions. Nonetheless, the average dependence on surfactant concentration shows a
decreasing repulsion as the amount of surfactant in the subphase increases (figure
\ref{fig:SDStime1}). A possible explanation of the time dependence of the repulsive interactions
is that the concentration of adsorbed SDS molecules on the oil-water interface changes with time,
which would change $\theta$. To rule out this possibility, the time dependence of the contact
angle on flat substrates was measured for the oil-water interface with 0.1mM and 0.5mM SDS added
to the aqueous phase. The contact angle was found to be constant for the case when 0.1mM was added
to the subphase over a 25 minute time period, while $\theta$ was found to fluctuate approximately
4$^\circ$ over a 75 minute time period when 0.5 mM SDS was added. However, fitting our constant
force measurements at $c_{SDS}$ = 0.5mM  using the Aveyard {\it et al.}\ model assuming a constant
oil dielectric constant requires a change in $\theta$ of more than 15$^\circ$ over one hour.  This
is larger than the range of change observed using direct contact angle experiments on planar
films.\\

The reasons for the slow time evolution in presence of SDS are not completely understood. The
process is too slow to be related to a diffusion phenomenon of ions across the surface. One
possibility is that surface hydrolysis of the SDS is taking place \cite{NAKA,Turner},  leading to
the formation of dodecan-1-ol from the hydrolysis of SDS. This will change the adsorption of SDS
onto the particles \cite{Turner} and may affect the local composition near the interface, which
has complex dynamics \cite{BENJ2}. In all, the experimental results show that the electrostatic
interaction becomes effectively more screened as time proceeds.

\subsubsection*{Effect of salt and SDS}

\begin{figure}[ht!]
\begin{center}
\includegraphics[width=13cm]{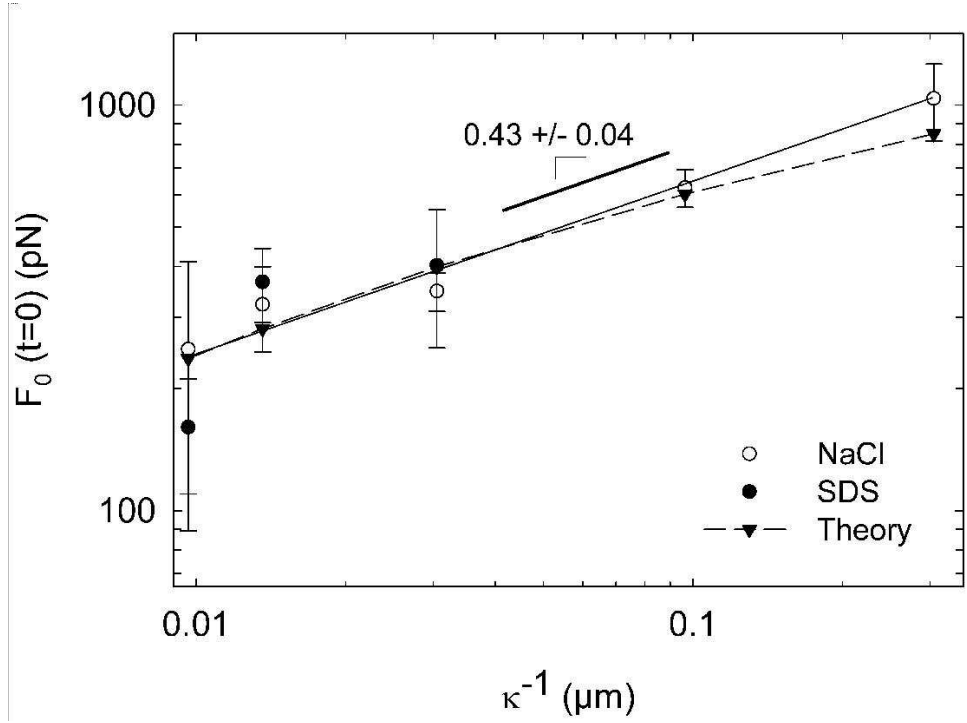}
{\caption{\small Effect of changing the Debye screening length on the interaction force for
systems containing NaCl or SDS. $F_0$ is the amplitude of the  dipole force parametrization
$F(r)=F_0\,(R/r)^4$.  The experimental results are compared with the expected dependence of the
charge renormalization theory (cf.\ equation \ref{eq:Oettel}) \cite{OettelCR}.  The solid line is
a power-law fit to the experimental data with a scaling exponent of 0.43$\pm$0.04.}
\label{fig:kappa}}
\end{center}
\end{figure}

Plotting the force after the addition of NaCl to the subphase, as a function of the Debye
screening length $\kappa^{-1}$, the predictions of the different theoretical approaches for the
electrostatic contribution with respect to the dependency on screening length can be evaluated.
The inverse screening length was calculated from
\begin{equation}
\kappa = \sqrt{\frac{1000 e^2 N_A}{\epsilon_W \epsilon_0 k_B T}2I} \qquad ,
\end{equation}
where I is the ionic strength of the electrolyte, $\epsilon_0$ is the permittivity of free space,
$k_B$ is the Boltzmann's constant, $T$ is the temperature, N$_A$ is Avogadro's Number and $e$ is
the elementary charge. For the pure water an ionic strength of 10$^{-6}$M was estimated, based on
conductivity measurements. Figure \ref{fig:kappa} shows the force as a function of $\kappa^{-1}$.
A weak, power law dependence of the magnitude of the force on the inverse screening length is
observed with an exponent of 0.43$\pm$0.04.  This can be compared to equation \ref{eq:Oettel} using $\kappa R$ and $\sigma^* = \sigma(e R)/k_B T \epsilon_0 \epsilon_w$, where $\sigma^*=7751$.  For a pre-factor of 15.8, the resulting predictions from the charge renormalization theory \cite{OettelCR} are in good agreement with the experimental values shown in figure \ref{fig:kappa}.
Interestingly, when treating SDS as a 1:1 electrolyte,
a very similar dependence is observed, suggesting that the primary effect of SDS is to act as an
electrolyte, although the change in the wetting properties could also play a significant role. As
the particles are pushed further into the oil phase, less charged surface area remains in contact
with the water.\\

\begin{figure}[ht!]
\begin{center}
\includegraphics[width=13cm]{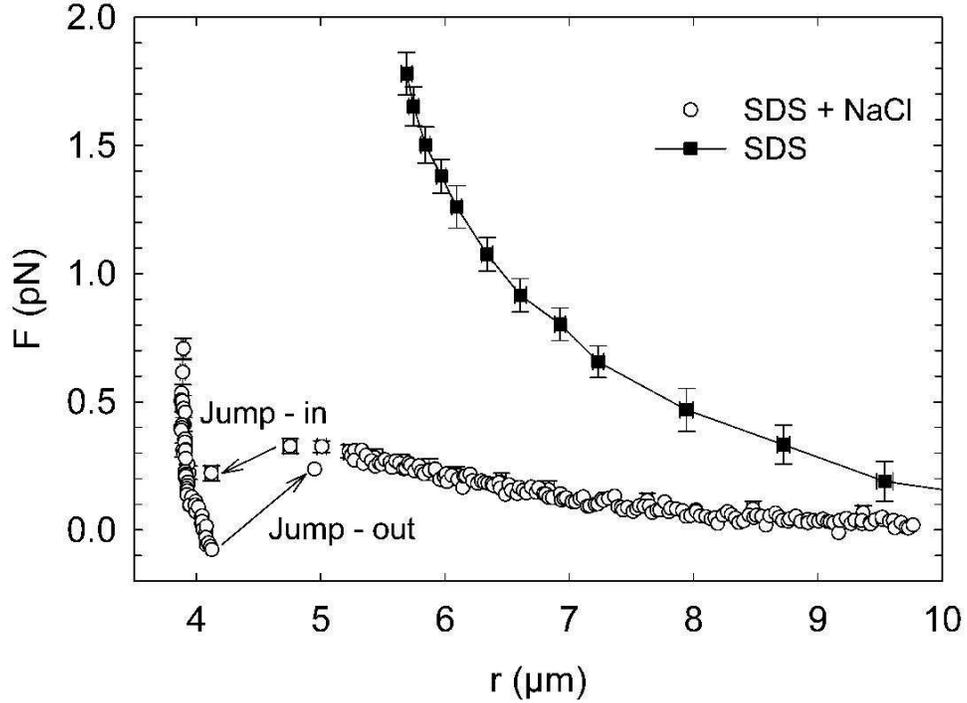}
{\caption{\small  Measurement of the pair interaction force
between between PS particles in 0.1mM SDS in the presence (open
symbols) and absence (closed symbols) of 0.25M NaCl 108 minutes
after spreading the particles.} \label{fig:SDStime3}}
\end{center}
\end{figure}

Finally, it should be noted that the {\sl combined} effect of salt and SDS on the interaction
between two particles seems to be more complex. In the presence of both SDS and salt, aggregation
is readily induced when particles are brought together, even at the lowest laser powers. For example,
figure \ref{fig:SDStime3} shows the force profile for the combination of 0.1 mM SDS and 0.25 M NaCl at 108 minutes after spreading the particles.
The range and magnitude of the repulsive
interaction is lower than measurements with SDS alone, and the power-law exhibits a weaker scaling exponent. Therefore, the addition of salt further
decreases the magnitude and scaling behavior of the far-field repulsion. In addition, the particles jump into a weak attractive minimum in the near-field, and exhibit a pull-out force with no hysteresis in the far-field interaction.  This observation rationalizes
earlier findings that combinations of salt and SDS lead to a more efficient destabilization of the
2D suspensions, whereas salt alone results in slow aggregation kinetics \cite{Reynaert}. \\

\subsection*{Anisotropy of the interactions}

\begin{figure}[ht!]
\begin{center}
\includegraphics[width=10cm]{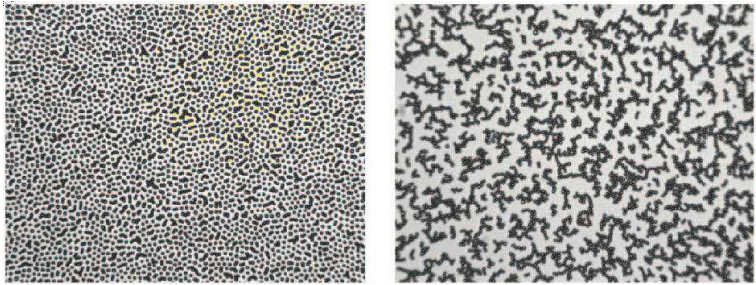} {\caption{\small Snapshots of the microstructure
during aggregation of an initially crystalline monolayer,
destabilized by the addition of 0.1 mM SDS  and 0.1 NaCl to the
aqueous subphase. Left: structure formed after the initial
aggregation. Right: structure formed after 1400 minutes. Scalebar
50 $\mu$m. \label{fig:aggregate}}}
\end{center}
\end{figure}

From detailed studies of the aggregation kinetics and suspension structure, several features
emerged that point to the role of anisotropic interactions \cite{Reynaert}. These features are
reproduced in aggregation experiments which correspond to a salt concentration of 0.1 M NaCl and
0.1 mM of SDS concentration, as discussed in the previous paragraph. Figure~\ref{fig:aggregate}
shows two snapshots during the aggregation process of a monolayer with an average surface area
coverage of approximately 0.30. The structures formed during the initial aggregation of the
monolayer are predominantly linear aggregates. However, near completion of the aggregation
process, the structure is rather dense. Typically, for systems with low SDS concentration in the
presence of NaCl, the fractal dimension $d_f$ is significantly higher than the one expected based
on the DLCA kinetics (i.e.\ in the range between 1.45 and 1.58 versus the expected value, 1.44).

\begin{figure}[ht!]
\begin{center}
\includegraphics[width=10cm]{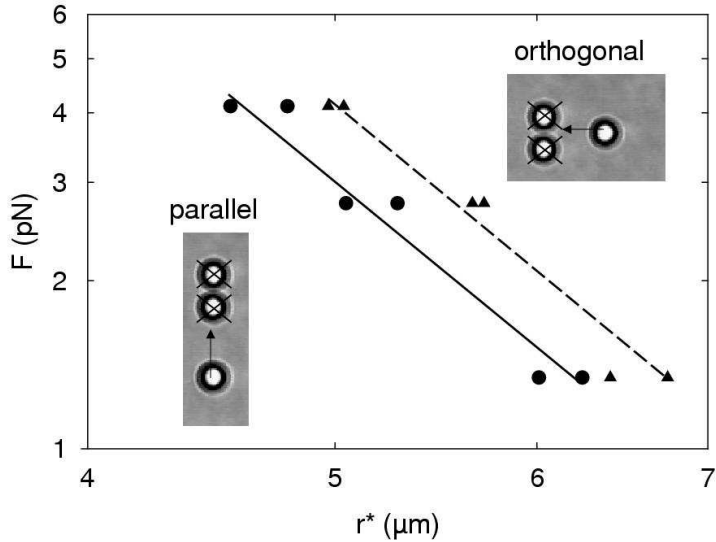}
{\caption{\small Measurements of the anisotropic interaction
between a particle and a dimer 45$\pm$10 minutes after the PS
particles were spread on the decane water interface with 0.5mM SDS
in the aqueous phase. (a) Comparison of the interaction along the
major axis of the dimer with a center-center separation of
3.5$\mu$m. (b) and the interaction along the minor axis of the
same dimer. \label{fig:aniso1}}}
\end{center}
\end{figure}

It is suggested that the linearity of the initial suspension structure upon aggregation is mainly
caused by the anisotropy of the electrostatic interactions between the initial doublets and
approaching single particles or doublets. To measure the magnitude of this interaction, different
particle configurations were created as shown in figure~\ref{fig:aniso1}. In the first
arrangement, two particles are held using strong stationary traps, at a separation short enough
that they form a dimer. A weak trap is then used to translate the third particle towards the dimer
along the axis connecting the particle centers. The same dimer arrangement is used for the second
experimental configuration, in which the third particle approaches the dimer orthogonal to the
connecting axis.

The results for the anisotropic repulsive interaction for a dimer are shown in
figure~\ref{fig:aniso1}.  In the case that the monomer approaches parallel to the dimer axis when
the dimer particle separation is 3.5$\mu$m, the interaction is found to follow a power scaling
with exponent $-4$, similar to the pair interaction. When approaching along the orthogonal axis a
similar power-law dependence is found, suggesting that in both cases, the electrostatic dipole
repulsion dominates. However, by comparing the prefactors from the power law fit of the
interaction, we find that the dimer/particle interaction along the orthogonal axis is a factor of
1.3 times greater than the particle pair interaction. This explains why particles in aggregating
system initially predominantly attach end-to-end, as the far-field energy barrier is smallest in
this direction, yielding the highest
probability for attachment. \\

The structures of the aggregating suspension, however, do not evolve linearly; they tend to
densify and exhibit fractal dimensions which can be significantly higher than expected from the
DLCA kinetics (1.47 to 1.58 depending on the compositional details the measured versus 1.44
expected). It has been argued previously that the anisotropy of the attractive capillary
interaction plays an important role in this case. This has already been calculated for the case of
gravitational induced capillary forces \cite{Krachlrev} or capillary forces induced by possible
surface roughness \cite{Stamou}. The curvature of the meniscus of two interacting particles has
two local minima occurring on either side of the doublet. The deformation in the interstitial
region can be complex, and depend on the wetting characteristics. However, in the present case the
electro-capillarity can be expected to be much more important as compared to the gravitational
forces \cite{OettelCR,Oettel1}. These electro-capillary interactions cannot be calculated assuming
a superposition of the interfacial profiles, but can be approximated using  a superposition of the
electric field \cite{Oettel}. The range and magnitude of the electro-capillary forces depend on
the charge density of the particles and the dielectric properties of the field. The effect of
changing medium composition on the electro-capillary interaction can be expected to be very
similar to the electrostatic repulsion and also the resulting spatial anisotropy would be similar
but different in sign. The electro-capillary interaction can only compete with the electrostatic
repulsion if $\epsilon_F = {\cal O}(1)$, see Eq.~(\ref{eq:Fcap}), i.e. if the total vertical
electrostatic force on the colloid is of the order of $\gamma R$. The contribution of possible
charges on the oil side to the total vertical electrostatic force is small whereas the
contribution of the charges on the water side can be large enough \cite{Oettel1}. It has also been suggested that a non-uniform surface charge distribution leads to dipolar attractions \cite{TONG}.
Finally, the Van
der Waals (VDW) attraction between particles will be affected by the wetting properties of the
particles. The Hamaker constant of decane is 25\% higher compared to water \cite{Israelachvili}.
Pushing the particles into the oil phase hence weakens the VDW forces somewhat. The nature and
anisotropy of these attractive interactions is not yet clear. Further work is required to
elucidate how the changing wetting and charge density properties affect these interactions,
possibly requiring an understanding of the multi-body effects.

\begin{figure}[ht!]
\begin{center}
\includegraphics[width=12cm]{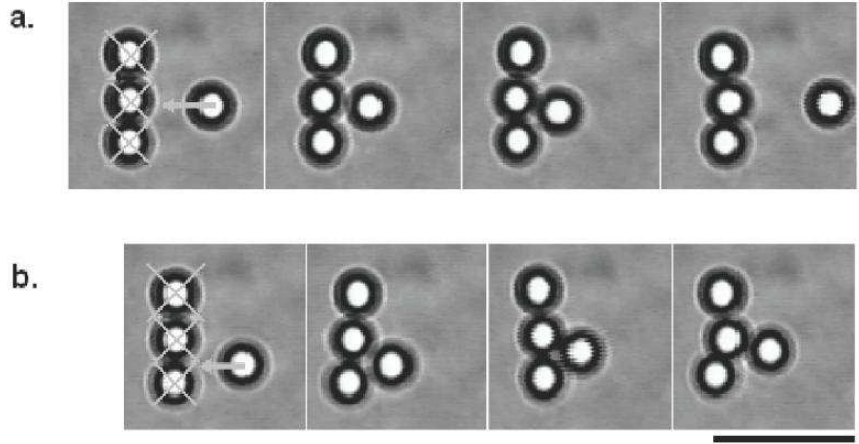} {\caption{\small Approach of a monomer along the
orthogonal to the axis of a trimer at the decane-water interface with 1.0mM SDS in the aqueous
subphase. (a) The monomer is initially perfectly aligned with the center particle of the trimer.
(b) This experiment was performed 1 minute later. The monomer is initially aligned with the region
between the center and bottom particles of the trimer. In both cases $F_{max}>20$pN.  The scale
bar is 10$\mu$m. Each time sequence goes from left to right.} \label{fig:aniso2}}
\end{center}
\end{figure}

In the present work, we evaluated experimentally whether the anisotropy in the attractive forces
could indeed be responsible for rearrangements that lead to a more complex structure with a
time-dependent aggregation. Using the optical traps, we assembled trimers two hours after
spreading the particles on the decane-water interface with 1.0mM SDS in the aqueous phase. The
maximum trapping force was approximately 20 pN. In figure \ref{fig:aniso2}a, the approaching
individual particle is perfectly aligned with the center particle of the trimer. We find that the
trimer bends away from the approaching individual particle, and at a close separation, the
individual particle begins moving downward, toward the region between the bottom two particles of
the trimer. However, the repulsive barrier to aggregation becomes greater than the maximum trap
strength, and the particle is released. At the same time, it appears that the triplet conformation
becomes bent towards the right, the direction of the monomer's approach. One minute later, the
approaching individual particle is aligned exactly between the two bottom particles of the trimer,
where the capillary attraction is maximum. We observe that the aggregate rotates within the traps
as the individual particle approaches due the repulsive interaction. However, the repulsive
barrier is reduced by the capillary force and the traps are now strong enough to overcome the
repulsion, and the individual particle aggregates to the center particle. In the system under
investigation the repulsive forces are still considerable and the overall anisotropy of the
interaction is dominated by the repulsion, hence when the optical trap holding the fourth particle
is removed, the particles reorient themselves to create a separation between the bottom particle
and the newly attached particle.

\section*{Conclusions}
Direct measurements of the pair-wise and multi-body interactions between polystyrene particles
spread at the decane-water interface are reported.  Using laser tweezers to measure the force
between particles, the effect of changing the Debye screening length leads to a power law
dependency of the interaction force. This suggests, contrary to earlier suggestions, that the
charges on the water side may still contribute significantly. The power law exponent of
0.43$\pm$0.04 is in good agreement with recent calculations, which account for charge
renormalization at the particle surface. However, the absolute magnitude of the measured force is
larger than what could be expected from only charges on the water side, hence it appears likely
that charges on both oil and water side contribute to the repulsion. It is notable that the form
of the repulsive interaction remains consistent with dipole--like behavior during the decrease of
its magnitude upon adding salt. Likewise, introduction of SDS in the aqueous sub-phase was shown
to result in a decrease in the repulsion between particles. Although treating SDS as a simple 1:1
electrolyte enables one to rationalize the observed initial decrease of the repulsion, the
steady-state values of the contact angle and surface tension are dependent on surfactant
concentration, and suggest an important role of the contact angle in controlling pair
interactions. By trapping multiple particles, the multi-body interactions were also investigated.
The stronger electrostatic interaction for single particles approaching orthogonal to an aggregate
explain the tendency for linear structures to form in the early aggregation process of 2D
suspensions.  It was also found that in the near-field, the capillary interactions in the vicinity
of aggregates are spatially anisotropic. As demonstrated in this work, the interactions between
particles at the water-oil interface in the presence of surfactants are significantly more
complex; however, this also presents a unique opportunity to tailor  the structure and rheology of
interfacial suspensions by controlling the structure of suspension through the wetting properties
of the particles and electrostatic interactions.  In turn, this enables one to tailor the
structure and stability of dispersed immiscible phases.

\small
\section*{Acknowledgements}
Basavaraj Madivala is acknowledged for his assistance in the gel trapping measurements. JV thanks
research council of the K.U. Leuven for financial support through GOA-2003/06, and a research
program of the Research Foundation-Flanders (FWO - Vlaanderen, project G.00469.05). EMF acknowledges support from NSF (CBET-0238689 and CBET-0553656).
MO acknowledges funding through the Collaborative Research Centre SFB-TR6 of the German Research
Council. This work was performed in the framework of a network of excellence SOFTCOMP (EU-6th
framework).

\newpage

\small\setlength{\baselineskip}{12pt}

\newpage
\subsection*{TABLE OF CONTENTS GRAPH}

\begin{figure}[ht!]
\begin{center}
\includegraphics[width=12cm]{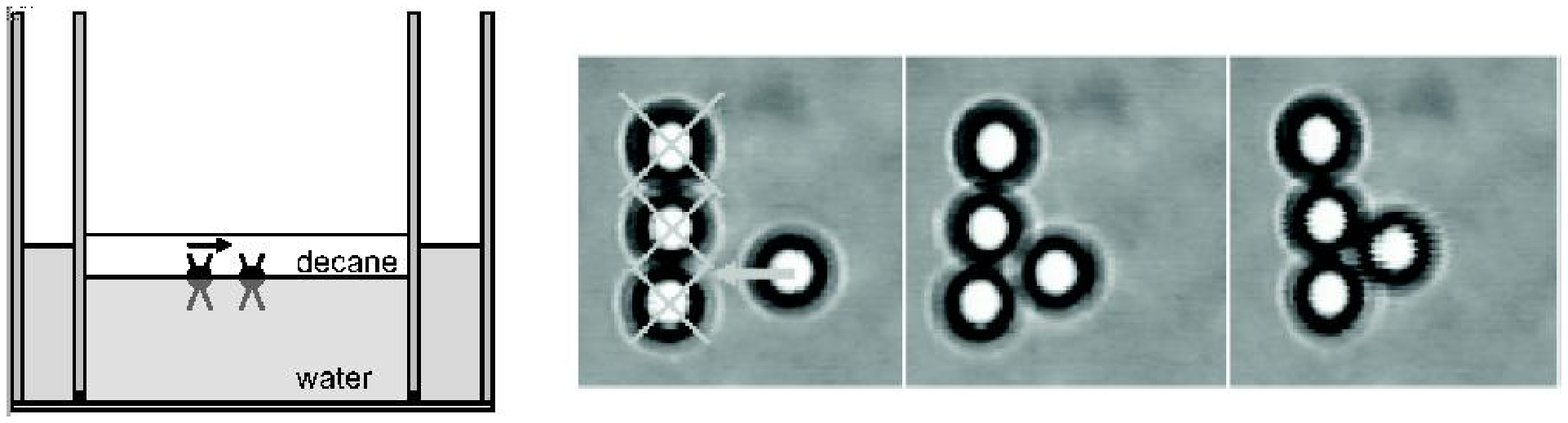}
\end{center}
\end{figure}

\end{document}